\documentclass[12pt]{article}
\usepackage{graphicx}

\newcommand\pubnumber{NuPhys2015-O'Sullivan}
\newcommand\pubdate{\today}

\def\duke{Department of Physics\\
Duke University, Durham, NC, USA}

\def\Title#1{\begin{center} {\Large #1 } \end{center}}
\def\Author#1{\begin{center}{ \sc #1} \end{center}}
\def\Address#1{\begin{center}{ \it #1} \end{center}}

\newcommand\pubblock{\rightline{\begin{tabular}{l} \pubnumber\\
         \pubdate  \end{tabular}}}
\newenvironment{Abstract}{\begin{quotation}  }{\end{quotation}}
\newenvironment{Presented}{\begin{quotation} \begin{center} 
             PRESENTED AT\end{center}\bigskip 
      \begin{center}\begin{large}}{\end{large}\end{center} \end{quotation}}

\def\beq{\begin{equation}}
\def\eeq#1{\label{#1}\end{equation}}
\def\eeqn{\end{equation}}

\def\beqa{\begin{eqnarray}}
\def\eeqa#1{\label{#1}\end{eqnarray}}
\def\eeqan{\end{eqnarray}}

\let\bar=\overbar

\def\Dslash{\not{\hbox{\kern-4pt $D$}}}
\def\dslash{\not{\hbox{\kern-2pt $\del$}}}

\def\msb{{\bar{\ssstyle M \kern -1pt S}}}

\begin{document}
\begin{titlepage}
\pubblock

\vfill
\Title{Atmospheric Neutrino Status}
\vfill
\Author{Erin O'Sullivan}
\Address{\duke}
\vfill
\begin{Abstract}

This conference proceeding discusses new results arising from atmospheric neutrino detection in the Super-Kamiokande and IceCube experiments. Super-Kamiokande has measured atmospheric neutrinos in the energy range of 100 MeV-10 TeV and uses this data set to conclusively measure the east-west effect to 8.0 (6.0) $\sigma$ for electron (muon)
neutrinos. IceCube is ideal for measuring high energy atmospheric neutrinos and has explored how different production channels for atmospheric neutrinos contribute to the total overall observed flux. The measurement is consistent with the conventional spectrum, produced by the decay of pions and kaon, while the contribution from the prompt channel (due to charm decay) is consistent with zero.  

\end{Abstract}
\vfill
\begin{Presented}
 NuPhys2015, Prospects in Neutrino Physics
 Barbican Centre, London, UK,  December 16--18, 2015
\end{Presented}
\vfill
\end{titlepage}
\def\thefootnote{\fnsymbol{footnote}}
\setcounter{footnote}{0}

\section{Introduction}

Atmospheric neutrinos are an abundant and naturally occurring source of neutrinos. These neutrinos are created when cosmic ray protons interact with nuclei in the Earth's atmosphere, producing pions or kaons that eventually decay into neutrinos either directly or via muons. Other channels for producing atmospheric neutrinos are also possible, including charm decay which can be a significant source for atmospheric neutrinos above 100 TeV. 

\begin{figure}[htb]
\centering
\includegraphics[height=3.0in, trim={0cm 5.0cm 0cm 5.0cm}, clip]{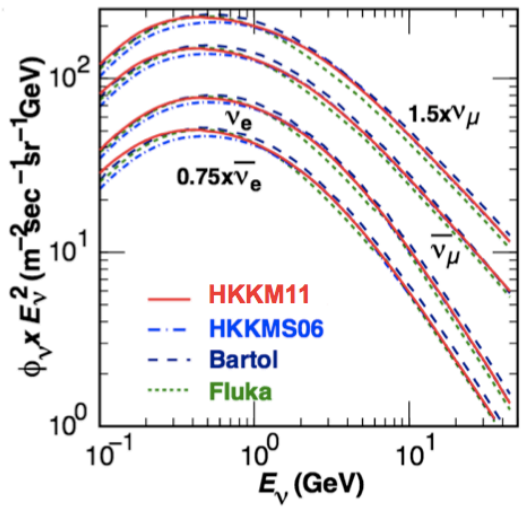}
\caption{Energy spectrum for the HKKM11 model for atmospheric
  neutrinos, compared with alternate models for atmospheric neutrino production (HKKMS06, Bartol, and Fluka). Figure modified from \cite{honda11}.}
\label{fig:atm_espec}
\end{figure}

Figure \ref{fig:atm_espec} shows the energy spectrum predicted by the HKKM11 model for atmospheric neutrinos. This figure shows the low energy region for atmospheric neutrinos, though in reality the neutrino energies span many orders of magnitude. 

This conference proceeding will focus on recent measurements of the
atmospheric neutrino flux from the Super-Kamiokande and IceCube
experiments. When combined, the data from these two experiments allows
us to probe a large portion of the atmospheric energy spectrum and will be compared with the HKKM11 model. The Super-Kamiokande
results will be used to examine predicted cosmic ray properties,
namely the east-west effect. The IceCube
results will be used to measure the contribution of the prompt
neutrino production channel on the overall atmospheric neutrino
spectrum. 

\section{Recent Results from Super-Kamiokande}

Super-Kamiokande (SK) is a 50 kton water-Cherenkov experiment located in Japan. The detector consists of an inner detector volume, which is viewed by 11,129 20'' photomultiplier tubes, and an optically separated outer detector, which is viewed by 1,885 8'' photomultiplier tubes.   

The data from SK are divided into four distinct phases. The SK-I phase includes data taken from 1996-2001 with the full 40\% photocoverage. SK-II refers to the data taken from 2002-2005 with reduced photocoverage, after an accident destroyed about half of the photomultiplier tubes. In the SK-III phase, between 2006-2008, data-taking with the full photocoverage resumed. Finally, the phase between 2008 and now is referred to as SK-IV and includes improvements to the electronics.   

SK has made many advancements in neutrino physics using the
atmospheric neutrino data set since it began taking data in 1996. In
1998, SK published a paper showing the first evidence of neutrino
oscillations \cite{fukuda98}, a result for which Takaaki Kajita won
the 2015 Nobel Prize in physics. In this paper, the authors showed
that there was a deficit in the number of neutrinos with a longer path
length (ie. those that were produced in the atmosphere on the other
side of the Earth), while the number of predicted neutrinos with a
shorter path length (ie. produced directly overhead) agreed with the expected number of events.

Since that seminal paper in 1998, SK has continued to produce results using the atmospheric neutrino data set. Tau appearance has been measured at the 3.8 $\sigma$ level in the SK atmospheric sample \cite{abe13}. Tests for exotic scenarios, such as Lorentz violation \cite{abe15a} and oscillations of atmospheric neutrinos into sterile neutrinos \cite{abe15b}, have been studied and world-leading limits were set. 

\begin{figure}[htb]
\centering
\includegraphics[height=2.5in]{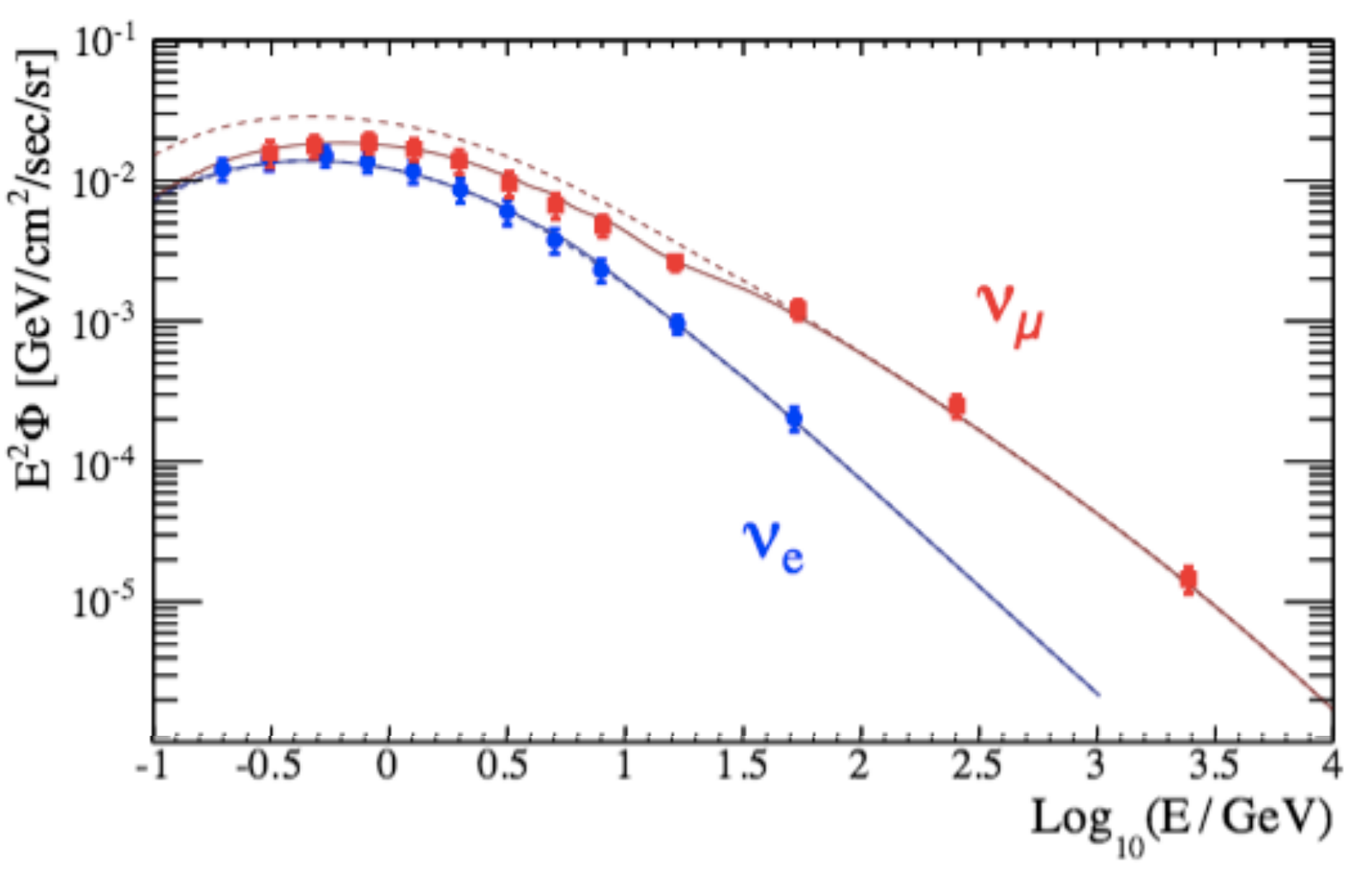}
\caption{Atmospheric neutrino fluxes as a function of energy for
  electron (blue) and muon (red) type neutrinos. The solid (dashed)
  line is the oscillated (unoscillated) prediction from the HKKM11 model. Figure modified from \cite{richards15}.}
\label{fig:SK_flux}
\end{figure}

Recently, SK released new atmospheric neutrino flux results
\cite{richards15}. These results show atmospheric neutrino fluxes from
neutrinos with energies spanning from 100 MeV - 10 TeV. Figure
\ref{fig:SK_flux} shows the SK results for neutrino flux as a function
of energy for electron and muon type neutrinos. The data fits well
with the oscillated HKKM11 model shown in
Figure \ref{fig:SK_flux}.    

\begin{figure}[htb]
\centering
\includegraphics[height=3.5in, trim={0cm 4.0cm 0cm 4.0cm}, clip]{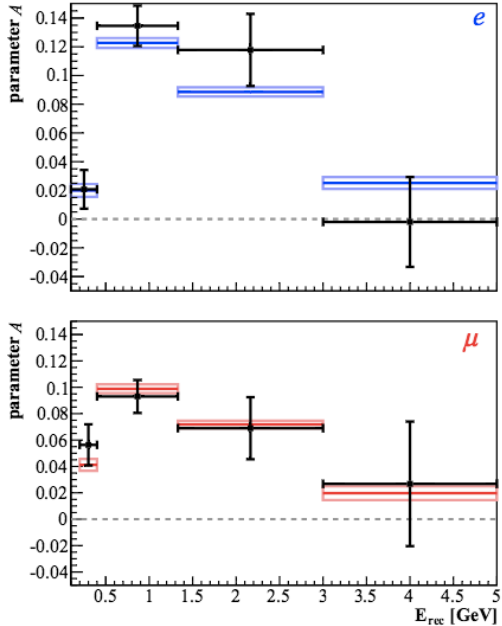}
\caption{Measurement of the east-west asymmetry, parameterized by
  A (as described in the text), as a function of reconstructed
  neutrino energy. The black points represent the SK data, while the boxes show the
  prediction from the SK Monte Carlo simulations. Figure modified from \cite{richards15}.}
\label{fig:SK_eastwest}
\end{figure}

The authors of \cite{richards15} also looked at the directional
asymmetry of the atmospheric neutrino flux. Super-Kamiokande has
previously measured a strong directional asymmetry (5 $\sigma$ for
electron neutrinos, and 2 $\sigma$ for muon neutrinos) \cite{futagami99}, know colloquially
as the east-west effect. This asymmetry arises from the deflection of
cosmic rays due to the Earth's magnetic field. The result is that more
cosmic rays (and thus more atmospheric neutrinos) should be coming
from the  west than from the east. In \cite{richards15}, this effect
was observed with a significance of 8.0 (6.0) $\sigma$ for electron (muon)
neutrinos. Figure \ref{fig:SK_eastwest} shows the measurement of the
east-west effect, parameterized by A, where

\begin{equation}
A=\frac{n_{east} - n_{west}}{n_{east} + n_{west}}
\end{equation}

and $n_{X}$ is the number of neutrinos coming from direction $X$
(where $X$ is broadly defined as east or west). 

As seen in Figure \ref{fig:SK_eastwest}, the east-west asymmetry
decreases as neutrinos increase in energy because they become
increasingly less affected by the Earth's magnetic field. In the lowest energy bin of Figure \ref{fig:SK_eastwest}, we also see a suppression of the east-west effect due to poor reconstruction of the incoming direction of the neutrino. 

\section{Recent Results from IceCube}

IceCube is a Cherenkov-based experiment that utilizes the
South Pole ice as a target for neutrino detection. The photosensors
that view the Cherenkov light, called digital optical modules (DOMs),
are arranged on strings suspended in the ice. Due to the large
volume of the detector, IceCube is well-suited to measure the rarer, highest
energy atmospheric neutrinos. 

As the energy of atmospheric neutrino increases, the flavour ratio of
the atmospheric neutrinos changes. At lower energies ($\sim$ 1 GeV),
the ratio of $(\nu_{\mu} +\bar{\nu_{\mu}})/(\nu_{e} +
\bar{\nu_{e}})$ is $\sim$ 2. As energy increases, muons no longer
decay in-flight into electron neutrinos and so this ratio becomes higher,
reaching $\sim$ 20 at 1 TeV. Furthermore, as the energy of the
incoming cosmic ray increases, the production mechanism of the atmospheric
neutrinos change. Atmospheric neutrinos below a few hundred GeV are
produced in what is known as the conventional spectrum, from the
decays of pions and kaons. Above that energy, however, the prompt
decay of charm mesons is thought to be a dominant production mechanism
for atmospheric neutrinos. 

\begin{figure}[htb]
\centering
\includegraphics[height=3.5in]{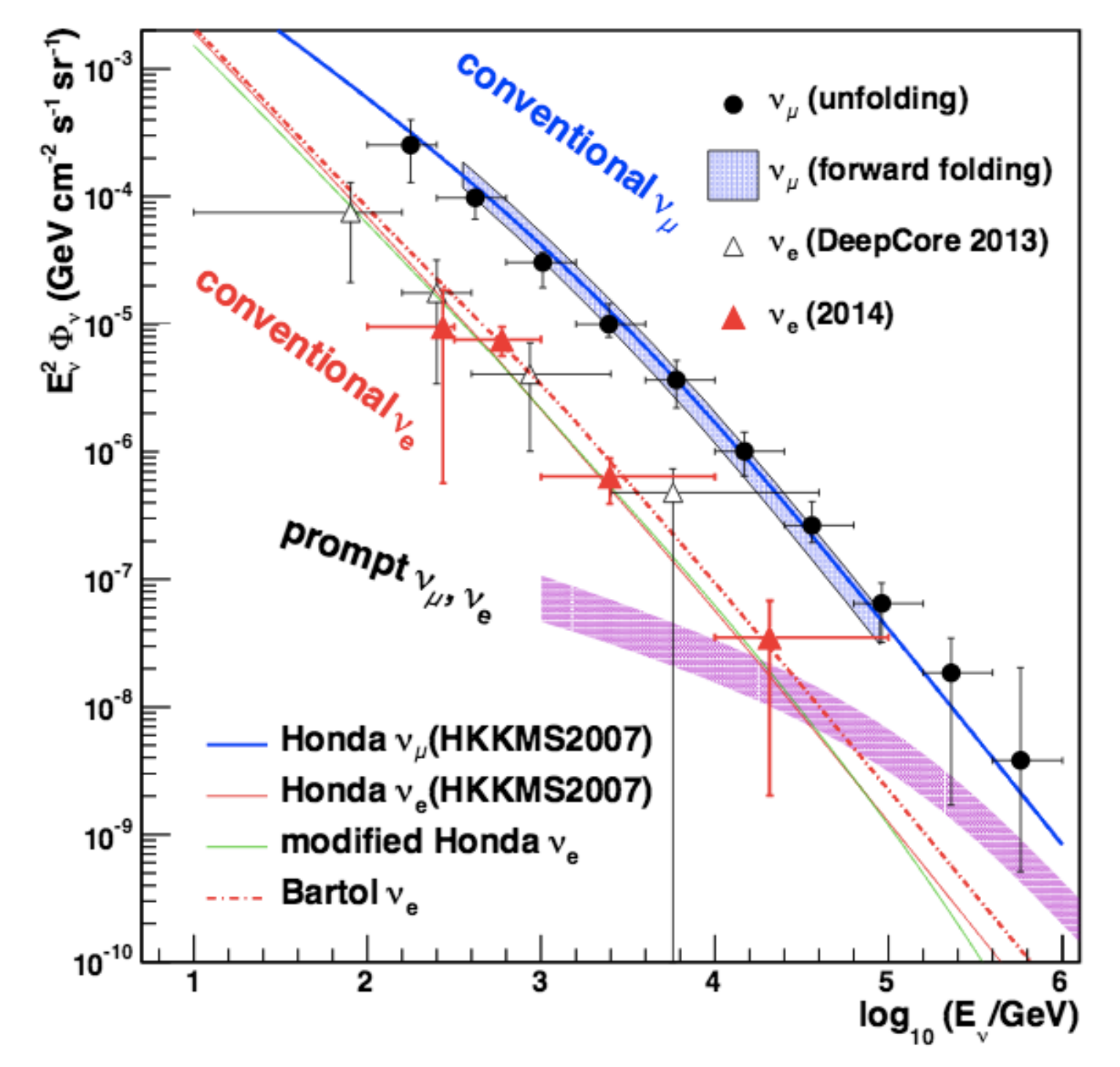}
\caption{IceCube atmospheric results, shown with the predicted conventional and
  prompt energy spectra.  Figure from \cite{aartsen15}.}
\label{fig:IC_prompt}
\end{figure}

In \cite{aartsen15}, IceCube set limits on the atmospheric neutrinos
created in the prompt decay mechanism. Figure \ref{fig:IC_prompt}
shows the IceCube data, along with the prediction for the conventional
and prompt spectrum. By calculating the normalization for the fit to
this data, limits can be set on the contributions from the
conventional and prompt production on the total
atmospheric neutrino spectrum. The authors of \cite{aartsen15} fit a
normalization of 0.0$^{+3.0}_{-0.0}$ to a modified ERS model
\cite{bhattacharya14}, meaning that no statistically significant
amount of atmospheric neutrinos are
measured to be from the prompt production channel. 

\section{SK and IceCube Combined Results}

\begin{figure}[htb]
\centering
\includegraphics[height=3.5in]{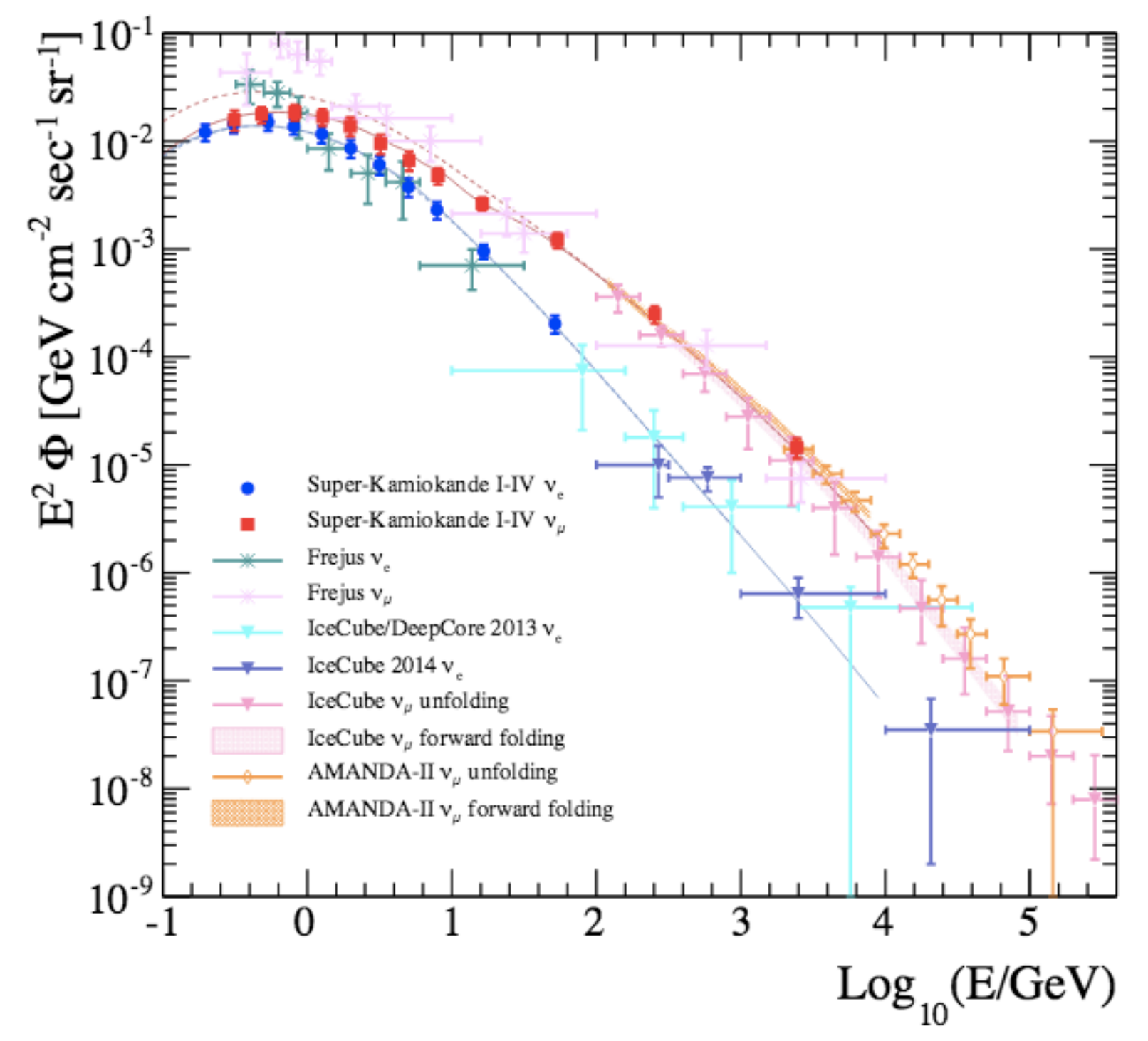}
\caption{Combined results from SK and IceCube, as well as Amanda-II
  and Frejus. The HKKM11 model is shown in solid (dashed) lines with (without) oscillations. Figure from \cite{richards15}.}
\label{fig:SK_IC_flux}
\end{figure}

Between the SK and IceCube results, atmospheric neutrinos are well
studied over the full energy range. Figure \ref{fig:SK_IC_flux} shows
the SK and IceCube flux measurements, as well as previous measurements from the 
Amanda-II \cite{abbasi10} and Frejus \cite{daum95}
experiments. Atmospheric neutrino flux measurements agree well with the
HKKM11 model over the entire energy range. 

\section{Future Atmospheric Neutrino Experiments}

In the coming years, new experiments will be constructed to further
study atmospheric neutrinos. 

The next-generation of the SK experiment,
Hyper-Kamiokande \cite{abe15HK}, is proposed to start data taking in the
mid-2020s. This water-Cherenkov detector would be 516 ktons, which is
an order of magnitude increase in total volume compared to SK. One of
the physics goals of Hyper-Kamiokande is to measure atmospheric neutrinos and use
these results to determine neutrino mass hierarchy, search for
CP-violation in the lepton sector (in combination with accelerator
neutrinos), and to improve the precision of the neutrino
mixing parameters. 

IceCube has proposed a future upgrade to their experiment whereby a
section of the volume is outfitted with DOM strings with closer
spacing, increasing the sensitivity of the experiment to lower energy
atmospheric neutrinos. This project, called Pingu \cite{aartsen14} , would enable
IceCube to measure atmospheric neutrinos down to 1 GeV.

\section{Conclusions}

Atmospheric neutrinos are good probes for neutrino properties, and can
also be used to study effects on the parent cosmic rays. This
conference proceeding focused on two atmospheric neutrino experiments:
Super-Kamiokande and IceCube. SK measures the lower energy region of
atmospheric neutrinos well due to the high photocoverage,
while the larger volume of IceCube enables precision high-energy
measurements. Current atmospheric neutrino models, such as HKKM11, are
in good agreement with the atmospheric neutrino data across the full neutrino
energy spectrum. Future generations of atmospheric neutrino
experiments, such as the proposed Hyper-Kamiokande and Pingu
experiments, will further utilize atmospheric neutrinos to discover
new and exciting effects in neutrino physics.

\end{document}